\documentclass[aps,prd,twocolumn,superscriptaddress,nofootinbib,10pt]{revtex4-2}
\usepackage{graphicx}
\usepackage{amssymb,amsmath,latexsym}
\usepackage{xcolor}
\usepackage[colorlinks=true,linktocpage=true,linkcolor=blue,citecolor=blue,urlcolor=blue]{hyperref}
\begin{document}
\title{Thermomagnetic enhancement of conserved-charge correlations in a lattice-constrained PNJL model}
\author{Bruno S. Lopes}
\email{bruno.lopes@acad.ufsm.br}
\affiliation{Departamento de Física, Universidade Federal de Santa Maria, 97105-900 Santa Maria, RS, Brazil}
\affiliation{Center for Nuclear Research, Department of Physics, Kent State University, Kent, OH 44242, USA}
\author{Arthur E. B. Pasqualotto} \email{arthur.pasqualotto@acad.ufsm.br}
\affiliation{Departamento de Física, Universidade Federal de Santa Maria, 97105-900 Santa Maria, RS, Brazil}
\author{Jandir J. Buttenbender Filho} \email{jandir.buttenbender@acad.ufsm.br}
\affiliation{Departamento de Física, Universidade Federal de Santa Maria, 97105-900 Santa Maria, RS, Brazil}
\author{Dyana C. Duarte}
\email{dyana.duarte@ufsm.br}
\affiliation{Departamento de Física, Universidade Federal de Santa Maria, 97105-900 Santa Maria, RS, Brazil}
\author{Ricardo L. S. Farias}
\email{ricardo.farias@ufsm.br}
\affiliation{Departamento de Física, Universidade Federal de Santa Maria, 97105-900 Santa Maria, RS, Brazil}
\affiliation{Center for Nuclear Research, Department of Physics, Kent State University, Kent, OH 44242, USA}
%
%
\begin{abstract}
Conserved-charge correlations offer a promising probe of magnetic fields in strongly interacting matter. We investigate their magnetic response and the finite-density equation of state within a lattice-constrained three-flavor Polyakov--Nambu--Jona-Lasinio model. A thermomagnetic scalar coupling, inferred by Gaussian process regression from lattice QCD results for the average light-quark condensate, incorporates inverse magnetic catalysis into the effective interaction. Without fitting conserved-charge observables, the resulting framework substantially improves their magnetic-field dependence and crossover behavior compared with the fixed-coupling model. In particular, it captures the pronounced magnetic enhancement of the baryon--electric-charge correlation and related susceptibility ratios, supporting their use as QCD magnetometers. The comparison demonstrates that the medium dependence of the effective interaction is essential to describe this response within the model. We further calculate pressure Taylor coefficients through sixth order and reconstruct the pressure and baryon density at nonzero baryon chemical potential. Our results show that lattice information from a single chiral observable constrains a broad thermodynamic response, connecting inverse magnetic catalysis, conserved-charge correlations, and the equation of state in a common framework. This provides a lattice-constrained basis for investigating magnetic-field signatures and finite-density thermodynamics in QCD matter.
\end{abstract}

\maketitle
%
\section{Introduction}~\label{sec:introduction}

The exploration of strongly interacting matter under extreme conditions of temperature and baryon density is one of the central goals of contemporary nuclear physics. Of particular interest is the structure of the quantum chromodynamics (QCD) phase diagram and the possible existence of a critical end point (CEP)~\cite{Asakawa:1989bq,Barducci:1989wi,Halasz:1998qr,Berges:1998rc,Stephanov:1998dy,Stephanov:1999zu}, which would mark the termination of a first-order chiral phase transition line and the onset of a smooth crossover. Determining whether such a critical point exists, and identifying its location in the $(T,\mu_B)$ plane, remains a major challenge for both theory and experiment~\cite{Stephanov:2024xkn}. 

On the theoretical side, effective field theories~\cite{Drischler:2021kxf}, functional methods~\cite{Herbst:2013ail,Fischer:2018sdj,Isserstedt:2019pgx,Fu:2019hdw,Gao:2020fbl,Gao:2020qsj,Dupuis:2020fhh,Fu:2022gou}, and lattice QCD calculations~\cite{Borsanyi:2010cj,Borsanyi:2013bia,HotQCD:2014kol,Ding:2015ona,Bazavov:2017dus,Ratti:2018ksb,Abbott:2023coj,Abbott:2024vhj,Borsanyi:2025dyp} have been extensively employed to investigate the phase structure of QCD, among other approaches~\cite{Hippert:2023bel,Chen:2024mmd,Shah:2024img,Ecker:2025vnb}. In recent years, fluctuation observables associated with conserved charges have emerged as particularly sensitive probes of critical phenomena, as higher-order cumulants and susceptibility ratios are expected to exhibit characteristic non-monotonic behavior in the vicinity of a CEP. In this context, important developments~\cite{Jeon:1999gr,Stephanov:2008qz,Stephanov:2009ra,Asakawa:2009aj,Friman:2011pf,Stephanov:2011pb,Bzdak:2012an,Morita:2012kt,Skokov:2012ds,Bzdak:2012ab,Ling:2015yau,Albright:2015uua,Vovchenko:2020gne,Vovchenko:2020kwg,Vovchenko:2020tsr,An:2021wof,Vovchenko:2022szk,Stephanov:2024mdj,Koch:2025cog,Braun-Munzinger:2026krf} have established direct connections between critical fluctuations and experimentally measurable cumulants of net-baryon number distributions. Experimentally, the search for the CEP was one of the main objectives of the Beam Energy Scan program at RHIC~\cite{STAR:2010vob,Bzdak:2019pkr,Du:2024wjm}. Measurements of higher-order cumulants of net-proton distributions by the STAR Collaboration have revealed intriguing non-monotonic trends as functions of the collision energy~\cite{STAR:2020tga,STAR:2025zdq}, motivating renewed interest in theoretical calculations of conserved-charge fluctuations and their relation to the QCD equation of state.

An additional ingredient naturally present in heavy-ion collisions is the generation of extremely intense magnetic fields~\cite{Voronyuk:2011jd,Bloczynski:2012en,STAR:2023jdd}, to which observation of the chiral magnetic effect~\cite{Kharzeev:2007jp,Fukushima:2008xe} is a highly pursued probe. Estimates indicate that non-central heavy-ion collisions can produce magnetic fields of the order of $eB \sim m_\pi^2$ or larger during the early stages of the collision~\cite{Skokov:2009qp,Bzdak:2011yy,Deng:2012pc,Shen:2025unr}. Magnetic fields thus influence chiral symmetry breaking~\cite{DElia:2012ems,Bali:2012zg,Mueller:2014tea}, modify the thermodynamic properties of QCD matter~\cite{Menezes:2008qt,Endrodi:2013cs}, and may affect both the location of the critical region and the behavior of fluctuation observables~\cite{Ferreira:2018pux,Vovchenko:2024wbg,Samanta:2025mrq}---see Refs.~\cite{Shovkovy:2012zn,Andersen:2014xxa,Miransky:2015ava,Bandyopadhyay:2020zte,Endrodi:2024cqn,Adhikari:2024bfa,Mustafa:2025uad} for thorough discussions on these topics. An important example is the phenomenon of magnetic ca\-ta\-lysis (MC), defined by an enhancement of the light-quark condensate with increasing magnetic field at zero temperature~\cite{Klevansky:1989vi,Gusynin:1994re,Gusynin:1994xp,DElia:2010abb}. Near the critical region, however, lattice QCD simulations at physical quark masses revealed the opposite trend, characterized by a suppression of the light-quark condensate and a decrease of the pseudocritical temperature with increasing field strength~\cite{Bali:2011qj,Bali:2012zg,Bruckmann:2013oba,Bali:2014kia}. This behavior is known as inverse magnetic catalysis (IMC), and is described in Ref.~\cite{Bruckmann:2013oba} as a back-reaction of sea quarks on gauge field configurations which dominates over valence quark interactions near the transition.

Although notoriously absent in traditional treatments based on effective models, IMC may be incorporated by the introduction of medium dependencies on the coupling strength~\cite{Farias:2014eca}. In standard effective chiral models, employing a constant scalar coupling leads exclusively to magnetic catalysis across all temperature regimes, failing to capture the chiral restoration behavior predicted by lattice QCD. This limitation stems primarily from the absence of explicit gluonic degrees of freedom, which prevents effective models from accounting for the previously mentioned back-reaction mechanism. A running thermomagnetic coupling $G(T, eB)$ can overcome this deficiency, and its behavior may be constrained directly by \textit{ab initio} lattice simulations. By incorporating this medium-dependent interaction strength, the effective coupling weakens near the transition region as the magnetic field intensity grows. Consequently, the chiral condensate and dynamical quark masses experience a suppression around the crossover temperature, naturally generating IMC and causing the pseudocritical temperature to decrease with increasing magnetic field. In addition to reconciling effective models with lattice QCD chiral phase diagrams, this approach provides a consistent foundation for evaluating thermodynamic observables and magnetic properties of quark matter, such as the renormalized magnetization. A more detailed analysis of these thermomagnetic parameterizations and their implications for the chiral phase transition can be found in Refs.~\cite{Farias:2014eca,Ferreira:2014kpa,Ayala:2014iba,Farias:2016gmy,Ahmad:2016iez,Yang:2017xdd,Avancini:2019wed,Tavares:2021fik,Farias:2022gaz,Islam:2023zpl,Tavares:2023oln,Coppola:2024uvz,Ding:2026tow}. Alternatively, the effective interactions may acquire medium dependencies through the introduction of nonlocal form factors, in turn reproducing IMC~\cite{Pagura:2016pwr,GomezDumm:2017iex}.

Intense magnetic fields may also induce significant effects on the QCD equation of state at finite density, altering nuclear matter properties and affecting stellar evolution~\cite{Most:2025kqf}. Theoretical investigations in this regime are more involved, as lattice QCD calculations are hindered by the sign problem~\cite{deForcrand:2009zkb}. Among alternative techniques developed to circumvent the issue, Taylor expansion schemes based on conserved-charge observables have been widely employed, especially due to the sensitivity of number fluctuations to the conjectured CEP. However, exploration of the QCD phase diagram is limited to low densities due to the nature of the expansion. Effective model descriptions thus remain important and may be further constrained by first-principles information on generalized susceptibilities and cumulant ratios from the lattice.

Several collaborations have recently computed the Taylor-expansion coefficients of the QCD pressure at finite density, temperature, and magnetic field~\cite{Borsanyi:2023buy,Ding:2023bft,Astrakhantsev:2024mat,Ding:2025jfz,Ding:2025nyh,Borsanyi:2026yna}. For very large magnetic fields, results from perturbation theory at two loops are also available~\cite{Fraga:2025juh}. Here, we confront these lattice calculations with a three-flavor Polyakov--Nambu--Jona-Lasinio (PNJL) model. A direct comparison, however, is challenging. Although standard NJL-type models naturally describe magnetic catalysis throughout the phase diagram, they fail to reproduce IMC near the crossover. Additionally, the resulting pseudocritical temperature is too high compared to lattice estimates, and the quark degrees of freedom remain excessively suppressed precisely in the region where conserved-charge fluctuations vary most rapidly. A realistic description of these observables therefore requires the effective interaction itself to encode the medium dependence associated with inverse magnetic catalysis, although its precise form remains a source of model uncertainty---see Refs.~\cite{Mao:2025toi,Mao:2026bfc} for investigations on second- and fourth-order correlations in different prescriptions. In our case, the issue is addressed by the introduction of a thermomagnetic coupling $G(T,eB)$, determined by a Gaussian process regression based on lattice results for the average light-quark condensate~\cite{Bali:2012zg}. This procedure is shown to be essential, since IMC strongly affects the region of interest around the transition.

The paper is organized as follows. In Sec.~\ref{sec:qns_definitions}, we briefly review the Taylor expansion scheme and the definitions of conserved-charge susceptibilities. Sec.~\ref{sec:su3_formalism} details the SU(3) PNJL model formalism, and Sec.~\ref{sec:parametrization} deals with its parametrization and determination of the thermomagnetic coupling. Numerical results are discussed in Sec.~\ref{sec:num_results}, where extensive comparisons with lattice QCD calculations are presented. Finally, Sec.~\ref{sec:conclusions} summarizes our conclusions and outlook.
%
\section{Susceptibilities and Taylor coefficients of the pressure}~\label{sec:qns_definitions}

At nonzero baryon density, direct lattice QCD simulations are hindered by the sign problem~\cite{deForcrand:2009zkb}. Thermodynamic information can nevertheless be obtained by expanding the dimensionless pressure around vanishing chemical potentials. We denote the scaled chemical potentials by $\hat\mu_X \equiv \mu_X/T$, where the index $X$ denotes the associated conserved quantity. Two equivalent bases are useful: the conserved-charge basis $\{B,Q,S\}$, associated with baryon number, electric charge, and strangeness, and the quark-flavor basis $\{u,d,s\}$. The pressure $P$ is thus written as
\begin{align}
\frac{P}{T^4} &= \sum_{i,j,k} \frac{1}{i!j!k!} \chi_{ijk}^{BQS}(\hat{\mu}_B)^i(\hat{\mu}_Q)^j(\hat{\mu}_S)^k, \label{eqn:BQSexp}\\
 &= \sum_{i,j,k} \frac{1}{i!j!k!} \chi_{ijk}^{uds}(\hat{\mu}_u)^i(\hat{\mu}_d)^j(\hat{\mu}_s)^k,\label{eqn:udsexp}
\end{align}
where the corresponding generalized susceptibilities are defined by
\begin{align}
\chi^{BQS}_{i,j,k} &= \frac{1}{VT^3} \left. \frac{\partial^{i+j+k} \ln Z}{(\partial \hat \mu_B)^i (\partial \hat \mu_Q)^j (\partial \hat \mu_S)^k} \right|_{\hat{\mu}_B = \hat{\mu}_Q = \hat{\mu}_S = 0} ,\\
\chi^{uds}_{i,j,k} &= \frac{1}{VT^3} \left. \frac{\partial^{i+j+k} \ln Z}{(\partial \hat \mu_u)^i (\partial \hat \mu_d)^j (\partial \hat \mu_s)^k} \right|_{\hat{\mu}_u = \hat{\mu}_d = \hat{\mu}_s = 0} ,
\end{align}
with $\dfrac{T}{V} \ln Z = P$, and $V$ being the volume of the system. The two bases are related by
\begin{align}
\mu_u &= \frac{1}{3}\mu_B + \frac{2}{3}\mu_Q \, , \nonumber\\
\mu_d &= \frac{1}{3}\mu_B - \frac{1}{3}\mu_Q \, , \nonumber\\
\mu_s &= \frac{1}{3}\mu_B - \frac{1}{3}\mu_Q - \mu_S \, .
\end{align}
At fixed temperature, the same linear relations hold for the normalized chemical potentials. The chain rule therefore gives
\begin{align}
\frac{\partial}{\partial \mu_B} &= \frac{1}{3} \frac{\partial}{\partial \mu_u} + \frac{1}{3} \frac{\partial}{\partial \mu_d} + \frac{1}{3} \frac{\partial}{\partial \mu_s} \, , \nonumber\\
\frac{\partial}{\partial \mu_Q} &= \frac{2}{3} \frac{\partial}{\partial \mu_u} - \frac{1}{3} \frac{\partial}{\partial \mu_d} - \frac{1}{3} \frac{\partial}{\partial \mu_s} \, , \nonumber\\
\frac{\partial}{\partial \mu_S} &= -\frac{\partial}{\partial \mu_s} \, .
\end{align}
Consequently, every conserved-charge susceptibility can be expressed as a linear combination of quark-flavor susceptibilities. For example, the first- and second-order baryon derivatives are
\begin{align}
\chi_{100}^{BQS} &= \frac{1}{3} \left( \chi_{100}^{uds} + \chi_{010}^{uds} + \chi_{001}^{uds} \right) \, , \nonumber\\
\chi_{200}^{BQS} &= \frac{1}{9} \left( \chi_{200}^{uds} + \chi_{020}^{uds} + \chi_{002}^{uds} + 2 \chi_{110}^{uds} \right. \nonumber\\
& \ \ \, \left.{} + 2 \chi_{101}^{uds} + 2 \chi_{011}^{uds} \right) \, .
\end{align}

For the comparison with Ref.~\cite{Astrakhantsev:2024mat}, we follow the trajectory $\mu_Q = 0$ and $\mu_S = \mu_B/3$. Through the relations above, this is equivalent to $\mu_u = \mu_d = \mu_B/3$ and $\mu_s = 0$, which approximately corresponds to zero-strangeness. Charge-conjugation symmetry implies that only even powers contribute at $\mu_B = 0$, so that
\begin{align}
\frac{P(\mu_B)}{T^4} &= c_0 + c_2 \hat{\mu}_B^2 + c_4 \hat{\mu}_B^4 + c_6 \hat{\mu}_B^6 + \mathcal{O}(\hat{\mu}_B^8) \, .
\end{align}
The coefficients $c_n \equiv c_n (T,eB)$ depend on the temperature and on the external magnetic field, and along this trajectory, are given by
\begin{align}
c_n &= \left[\frac{1}{n!}\left(\frac{\partial}{\partial \hat{\mu}_B} + \frac{1}{3} \frac{\partial}{\partial \hat{\mu}_S}\right)^n \frac{P}{T^4}\right] \Bigg|_{\hat{\mu}_B = \hat{\mu}_S = 0} .
\end{align}
Alternatively, setting $\mu_S = \mu_B/3$ before taking any derivatives, they may be simply expressed as $c_n = (1/n!) \, (d / d\hat{\mu}_B)^n (P/T^4) |_{\hat{\mu}_B = 0}$ along the specified trajectory.

The other finite-density observables shown below follow directly from this expansion. Up to sixth order, we have
\begin{align}
\frac{\Delta p}{T^4} &= c_2 \hat{\mu}_B^2 + c_4 \hat{\mu}_B^4 + c_6 \hat{\mu}_B^6 \, , \nonumber\\
\frac{n_B}{T^3} &= 2c_2 \hat{\mu}_B + 4c_4 \hat{\mu}_B^3 + 6c_6 \hat{\mu}_B^5 \, ,
\label{eqn:observables_expansion}
\end{align}
where $\Delta p \equiv P(T,eB,\mu_B) - P(T,eB,0)$. These expressions also show why the baryon number density is significantly more sensitive to $c_6$ than the pressure.

For notational simplicity, we use $\chi_2^B \equiv \chi_{200}^{BQS}$ and analogous shorthand notation for the other diagonal and mixed second-order susceptibilities. In the following, we introduce the effective-model framework used to evaluate these quantities.
%
\section{Three-flavor PNJL model formalism}~\label{sec:su3_formalism}

We employ the three-flavor Polyakov--Nambu--Jona-Lasinio model for the light $u$, $d$, and $s$ quark sectors, collected in the flavor multiplet $\psi=(\psi_u,\psi_d,\psi_s)^T$. The Polyakov loop provides a statistical description of confinement~\cite{Fukushima:2003fw,Ratti:2005jh}, while the local fermionic interactions account for spontaneous chiral-symmetry breaking and flavor mixing~\cite{Nambu:1961tp,Nambu:1961fr,Klevansky:1992qe,Hatsuda:1994pi,Buballa:2003qv}. The three-flavor PNJL Lagrangian in an external electromagnetic field is given by
\begin{align}
\mathcal{L} &= \bar{\psi} \left(i \gamma_\mu D^\mu - \hat m \right) \psi - \frac{1}{4} F^{\mu\nu} F_{\mu\nu} + \nonumber\\
&+ \mathcal{L}_\mathrm{sym} + \mathcal{L}_\mathrm{KMT} - \mathcal{U}(\Phi,\bar{\Phi}) \, .
\end{align}
Here, $\hat{m} = \mathrm{diag}(m_u,m_d,m_s)$ is the current-quark mass matrix, $F^{\mu\nu} = \partial^\mu A^\nu - \partial^\nu A^\mu$ the electromagnetic field tensor and $D^\mu = \partial^\mu + i Q A^\mu - i\delta_{0}^{\mu} \mathcal{A}^0$, $A^\mu$ being the electromagnetic four-potential and $\mathcal{A}^\mu = \delta_{0}^{\mu} \mathcal{A}^0$ the temporal background gauge field which incorporates Polyakov-loop dynamics~\cite{Ratti:2005jh}. The charge matrix is $Q = \mathrm{diag}(2e/3,-e/3,-e/3)$, with $e = \sqrt{4\pi\alpha}$ and $\alpha \simeq 1/137$. The chirally symmetric four-fermion interaction $\mathcal{L}_{\mathrm{sym}}$ and the six-fermion Kobayashi--Maskawa--'t~Hooft determinant $\mathcal{L}_{\mathrm{KMT}}$ are
\begin{align}
\mathcal{L}_\mathrm{sym} &= G \sum^8_{a=0} \left[(\bar\psi \lambda_a \psi)^2 + (\bar\psi i \gamma_5\lambda_a \psi)^2\right] \, , \\
\mathcal{L}_\mathrm{KMT} &= -K \left\{\det\left[\bar \psi(1 + \gamma_5)\psi\right] + \right. \nonumber \\
&\left. + \det\left[\bar \psi(1 - \gamma_5)\psi\right]\right\} \, .
\end{align}
The matrices $\lambda_a$ ($a=1,\ldots,8$) are the Gell-Mann generators, and $\lambda_0 = \sqrt{2/3} \, \mathbb{I}_{3\times3}$ ($\mathbb{I}$ being the identity) completes the flavor basis with the normalization $\mathrm{Tr}(\lambda_a\lambda_b)=2\delta_{ab}$. The determinants in $\mathcal{L}_{\mathrm{KMT}}$ are taken in flavor space and explicitly break $U(1)_A$, thereby generating the required flavor mixing in the pseudoscalar singlet sector. For a static and homogeneous magnetic field $\mathbf B = B \hat{\mathbf z}$, we use the Landau gauge $A^\mu = (0,0,xB,0)$. Finally, $\mathcal{U}(\Phi,\bar{\Phi})$ is the effective potential associated with the Polyakov loop
\begin{align}
\Phi &= \frac{1}{N_c} \left\langle \mathrm{Tr}_c \, \mathcal{P} \exp \left( i\int_{0}^{\beta} d\tau \mathcal{A}_4 \right) \right\rangle \, ,
\end{align}
with $\bar{\Phi}$ being its conjugate, $\mathcal{A}_4 = - i \mathcal{A}_0$, $\mathrm{Tr}_c$ the trace over color space and $N_c = 3$ the number of colors. We work in the Polyakov gauge, in which $\mathcal{A}_4$ is static and diagonal~\cite{Fukushima:2003fw}. The parametrization of $\mathcal{U}(\Phi,\bar{\Phi})$ is addressed further below.

In the mean-field approximation, the matter contribution to the grand potential is evaluated using magnetic-field-independent regularization (MFIR)~\cite{Ebert:1999ht,Ebert:2003yk,Allen:2015paa}.\footnote{The designation ``Magnetic Field Independent Regularization'' was introduced in Ref.~\cite{Allen:2015paa}, while the separation of magnetic-field-dependent finite contributions from vacuum divergences in NJL-type models goes back to Refs.~\cite{Ebert:1999ht,Ebert:2003yk}. More applications and discussions can be found in Refs.~\cite{Menezes:2008qt,Duarte:2015ppa,Farias:2016gmy,Avancini:2019wed,Avancini:2020xqe}.} We omit the classical Maxwell contribution $B^2/2$, which is independent of the quark chemical potentials and is not included in the matter pressure used for comparison with lattice QCD. The resulting potential reads:
\begin{align}
\Omega = \mathcal{U}(\Phi, \bar{\Phi}) + \sum_f \omega_f + 2G \sum_f \phi_f^2 - 4 K \phi_u \phi_d \phi_s \, .
\end{align}
The sums run over flavors $f \in \{u,d,s\}$, $\phi_f = \langle \bar{\psi}_f \psi_f \rangle$ are the chiral condensates, and $\omega_f$ denotes the corresponding quasiparticles contribution. Within MFIR, both quantities are separated into a regularized vacuum term and finite magnetic and medium terms:
\begin{align}
\omega_f = \omega_f^{\text{vac}} + \omega_f^{\text{mag}} + \omega_f^{\text{med}} \, ,
\end{align}
and
\begin{align}
\phi_f = \phi_f^{\text{vac}} + \phi_f^{\text{mag}} + \phi_f^{\text{med}} \, .
\end{align}
The vacuum, purely magnetic, and thermal-medium contributions are, respectively,
\begin{align}
\omega_f^{\rm vac} &= \frac{N_c}{8\pi^2} \left[ M_f^4 \log\left( \frac{\Lambda + \epsilon_f}{M_f} \right) - \epsilon_f \Lambda (\Lambda^2 + \epsilon_f^2) \right] \, , \\
\omega_f^{\rm mag} &= - \frac{N_c (|q_f B|)^2}{2\pi^2} \left[ \zeta^\prime(-1,x_f) + \frac{1}{4}x_f^2 \right. \nonumber\\
&\left.- \frac{1}{2}(x_f^2 - x_f)\log x_f \right] \, , \\
\omega_f^{\rm med} &= - T \sum_{k = 0}^{+\infty} (2 - \delta_{0k}) \frac{|q_f B|}{2\pi^2} \nonumber \\
&\times \int_{0}^{+\infty} dp \, \left[ \log (z_f^{-}) + \log (z_f^{+}) \right] \,,
\end{align}
with the definitions
\begin{align}
x_f &= \frac{M_f^2}{2|q_f B|} \, , \nonumber\\
\epsilon_f &= (\Lambda^2 + M_f^2)^{1/2} \, , \nonumber\\
E_f &= (p^2 + M_f^2 + 2 |q_f B| k)^{1/2} \, ,
\end{align}
where $k$ labels the Landau levels, $2-\delta_{0k}$ accounts for the spin degeneracy, $\beta = 1/T$, and $\zeta'(s,x) = \partial \zeta(s,x) / \partial s$ denotes the derivative of the Hurwitz zeta function. The thermal factors are modified by the Polyakov loop and are given by
\begin{align}
z_f^- &= 1 + 3 \Phi e^{-\beta(E_f - \mu_f)} + 3 \bar{\Phi} e^{-2\beta(E_f - \mu_f)} \nonumber\\
&+ e^{-3\beta(E_f - \mu_f)} \, , \\
z_f^+ &= 1 + 3 \bar{\Phi} e^{-\beta(E_f + \mu_f)} + 3 \Phi e^{-2\beta(E_f + \mu_f)} \nonumber\\
&+ e^{-3\beta(E_f + \mu_f)} \, .
\end{align}

The corresponding contributions to the chiral condensates are
\begin{align}
\phi_f^{\rm vac} &= - \frac{N_c M_f}{2\pi^2} \left[ \Lambda \epsilon_f - M_f^2 \log \left( \frac{\Lambda + \epsilon_f}{M_f} \right) \right] \, , \\
\phi_f^{\rm mag} &= - \frac{N_c M_f |q_f B|}{2\pi^2} \left[ \log \Gamma(x_f) - \frac{1}{2} \log(2\pi) + x_f \right.\nonumber\\
&\left.- \frac{1}{2} (2x_f - 1) \log x_f \right] \, , \\
\phi_f^{\rm med} &= \sum_{k = 0}^{+\infty} (2 - \delta_{0k}) \frac{N_c M_f |q_f B|}{2\pi^2} \nonumber\\
&\times \int_{0}^{+\infty} dp \, \frac{1}{E_f} \left( n_f^- + n_f^+ \right) \, ,
\end{align}
with the Polyakov-modified occupation numbers
\begin{align}
n_f^- &= \frac{ \Phi e^{-\beta(E_f - \mu_f)} + 2 \bar{\Phi} e^{-2\beta(E_f - \mu_f)} + e^{-3\beta(E_f - \mu_f)} }{z_f^-} \, , \nonumber\\
n_f^+ &= \frac{ \bar{\Phi} e^{-\beta(E_f + \mu_f)} + 2 \Phi e^{-2\beta(E_f + \mu_f)} + e^{-3\beta(E_f + \mu_f)} }{z_f^+} \, .
\end{align}

The constituent quark masses are determined self-consistently from
\begin{align}
 M_i = m_i - 4G \phi_i + 2K \phi_j \phi_k \, ,
\end{align}
with $(i,j,k)$ being cyclic permutations in $(u,d,s)$. At each $(T,eB,\{\mu_f\})$, these gap equations are solved together with the stationary conditions for the Polyakov loop:
\begin{align}
\frac{\partial \Omega}{\partial \Phi} = 0 \, , \qquad \frac{\partial \Omega}{\partial \bar{\Phi}} = 0 \, .
\end{align}
At real chemical potential, $\Phi$ and $\bar{\Phi}$ are treated as independent real mean fields. When using the thermomagnetic coupling introduced in Sec.~\ref{sec:parametrization}, we evaluate $G(T,eB)$ at given values of $T$ and $eB$ before solving this coupled system.

To close the mean-field system, we adopt the logarithmic Polyakov-loop potential of Ref.~\cite{Roessner:2006xn}; alternative parametrizations are reviewed in Ref.~\cite{Fukushima:2017csk}:
\begin{align}
\frac{\mathcal{U} (\Phi, \bar{\Phi})}{T^4} &= - \frac{a(T)}{2} \bar{\Phi} \Phi + b(T) \log \left[ 1 - 6\bar{\Phi}\Phi \right.\nonumber\\
&\left. + 4(\bar{\Phi}^3 + \Phi^3) - 3(\bar{\Phi}\Phi)^2 \right] \, ,
\end{align}
where
\begin{align}
a(T) &= a_0 + a_1 \left( \frac{T_0}{T} \right) + a_2 \left( \frac{T_0}{T} \right)^2 \, , \nonumber\\
b(T) &= b_3 \left( \frac{T_0}{T} \right)^3 \, .
\end{align}
The coefficient values are $a_0 = 3.51$, $a_1 = -2.47$, $a_2 = 15.2$, and $b_3 = -1.75$. In the pure-gauge parametrization, $T_0 = 270~\mathrm{MeV}$ reproduces the deconfinement transition of Yang--Mills theory. The presence of dynamical quarks lowers this scale~\cite{Schaefer:2009ui,Carlomagno:2021gcy}, and throughout the present calculation we adopt $T_0 = 210~\mathrm{MeV}$.

\section{Model parametrization and coupling regression}~\label{sec:parametrization}

In this section, we discuss the model vacuum parametrization procedure and the regression model used to determine the coupling $G$ as a function of both the temperature and magnetic field.
%
\subsection{Vacuum parametrization}

The non-renormalizable nature of the NJL model requires the introduction of an ultraviolet regulator, which cannot be removed through the usual renormalization procedure and therefore becomes a physical parameter of the model. The standard parametrization procedure for the three-flavor NJL model was widely discussed in the literature~\cite{Klevansky:1992qe,Hatsuda:1994pi,Rehberg:1995kh,Buballa:2003qv,Kohyama:2016fif,Costa:2019bua,CamaraPereira:2020rtu}, and a detailed and self-contained presentation of the equations required for the model parametrization can be found in Ref.~\cite{Lopes:2021tro}.

Here, we employ the usual three-dimensional cutoff $\Lambda$ on the upper limit of divergent momentum integrals. The SU(3)$_f$ NJL model is parametrized in vacuum by fitting the cutoff $\Lambda$, the four-fermion coupling $G$, the six-fermion coupling $K$, and the strange current quark mass $m_s$ to reproduce the physical values of the pion mass $m_\pi$, the pion decay constant $f_\pi$, the kaon mass $m_K$, and the eta-prime mass $m_{\eta'}$. In this work, we matched the value of $f_\pi$ to that of Ref.~\cite{Bali:2012zg}, in order to consistently fit the values of the thermomagnetic coupling $G(T,eB)$ to the chiral condensate data therein. This regression procedure is described in the next subsection. For clarity, we denote the vacuum-parametrized value of the coupling as $G_0$. The numerical values of input and output parameters in the vacuum parametrization are given in Tab.~\ref{tab:vac_parameters}.
\begin{table}[htbp]
\centering
\caption{Parameter set used in the present study.}
\small
\begin{ruledtabular}
\begin{tabular}{cc}
Input parameters & Output parameters \\
\hline
\begin{tabular}{@{}c@{}}
$m_u = m_d = 5$ MeV\\
$f_\pi = 86.0$ MeV \\
$m_\pi = 135.0$ MeV \\
$m_K = 497.7$ MeV \\
$m_{\eta'} = 960.8$ MeV
\end{tabular}
& 
\begin{tabular}{@{}c@{}}
$\Lambda = 648.74$ MeV \\
$G_0 = 3.40$ GeV$^{-2}$\\
$K = 133.57$ GeV$^{-5}$ \\
$m_s = 172.65$ MeV
\end{tabular} \\
\end{tabular}
\end{ruledtabular}
\label{tab:vac_parameters}
\end{table}
%
\subsection{Gaussian process regression for thermomagnetic effective coupling}

As part of our model in the finite temperature and magnetic field regime, we impose a thermomagnetic-dependent coupling $G(T,eB)$
such that the values of the average light-quark condensate $\bar\Sigma(T,eB) = (\Sigma_u + \Sigma_d)/2$ from the lattice are reproduced within some degree of uncertainty. The input for our model comes from results obtained in Ref.~\cite{Bali:2012zg}, with $\Sigma_f$ defined as
\begin{align}
    \Sigma_f(T,eB) = 1 - \frac{2 m_f}{m_\pi^2 f_\pi^2}\left[\phi_f(T,eB) -\phi_f(0,0) \right] \, .
\end{align}
This additional constraint is solved self-consistently along with the PNJL model gap equations, providing $G$ with a medium dependence.
From this procedure, we acquired a discrete dataset $\{T_i,eB_i,G_i,{\sigma_G}_i\}$, where the standard deviation of $G$, $\sigma_{G_i}$, was computed through simple error propagation.

To obtain a smooth function $G(T,eB)$ from this discrete dataset, we trained a two-dimensional Gaussian Process (GP) regression model~\cite{edward2005rasmussen}.
A GP regression model generates predictions for the value of a function based on the available data using Bayesian inference, requiring a prior and a covariance function, often known as a kernel, as inputs.
Here we use the squared exponential kernel, which requires one hyperparameter for each dimension of the independent variables, corresponding to the correlation lengths ${l_T,l_{eB}}$. We choose $l_T = 0.035$ and $l_{eB} = 0.2$.
The correlation lengths weigh the influence of the data on the predicted values as a function of the distance in the parameter space $(T,eB)$ through the kernel. 
The prior for $G(T,eB)$ was chosen to have a Woods--Saxon-type form:
\begin{align}
     G_{\text{prior}}(T,eB) = \frac{\kappa}{(1+e^{(T-\eta)/\delta}) (1 + \gamma eB)},
\end{align}
with $\kappa = 3.40~\mathrm{GeV}^{-2}$, $\eta = 0.150~\mathrm{GeV}$, $\delta = 0.025~\mathrm{GeV}$ and $\gamma = 1~\mathrm{GeV}^{-2}$.
The GP regression model for $G(T,eB)$ was implemented in Julia Language~\cite{bezanson2017julia} with the \texttt{GaussianProcesses.jl} library~\cite{fairbrother2022gaussianprocesses}.
\begin{figure}[t]
    \centering
    \includegraphics[width=0.99\linewidth]{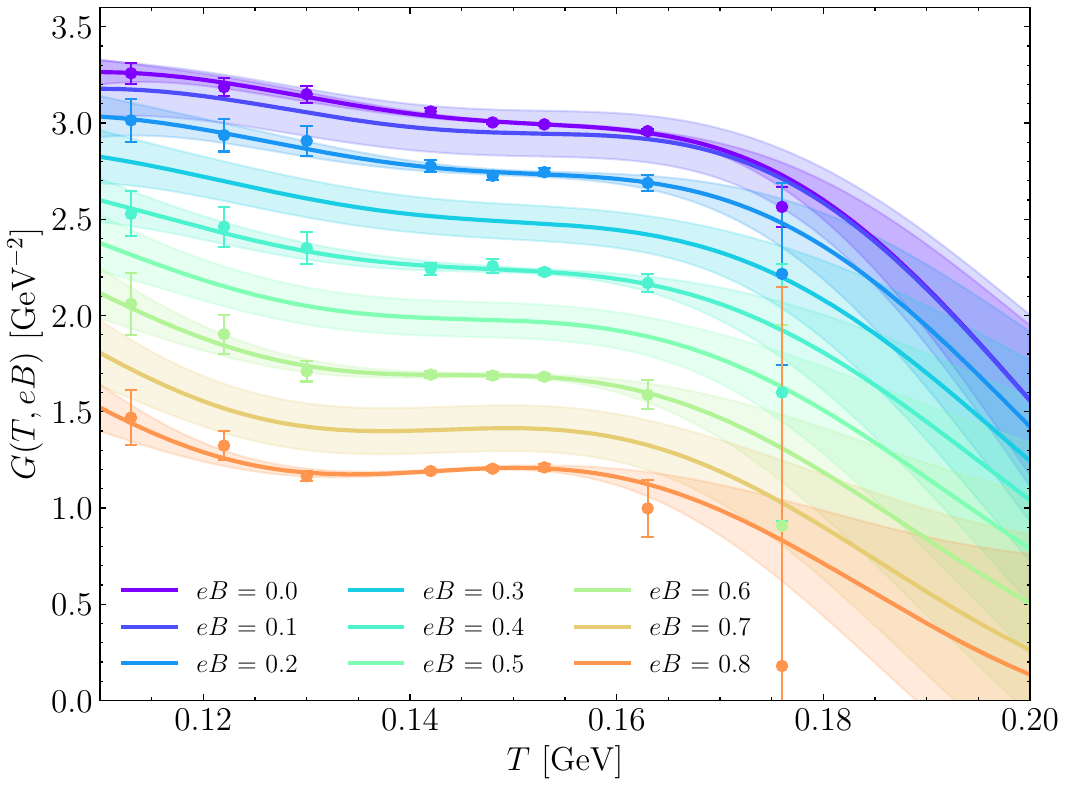}
    \caption{Values of $G(T,eB)$ as functions of $T$ for different values of $eB$ given in GeV$^2$. The solid lines correspond to the mean value of $G(T,eB)$ predicted by the GP, and the bands around each line represent the standard deviation of the prediction. The data points and respective error bars are the values used in the GP regression, obtained by fitting the light-quark condensate to the lattice data from Ref.~\cite{Bali:2012zg}.}
    \label{fig:G_predicted_values}
\end{figure}

The predicted mean values of $G(T,eB)$ as a function of $T$ for different slices of $eB$ are shown in Fig.~\ref{fig:G_predicted_values}. The colored regions around each curve indicate the one-standard-deviation uncertainty band of $G(T,eB)$ predicted by the GP regression model. The data points and error bars correspond to the input data used for training. As can be seen, the uncertainty band is wider for the magnetic-field slices where no lattice QCD data are available. This is characteristic of GP regression models: predicted values of the resulting set of functions follow a covariance law. In this case, the covariance law is a function of the Euclidean distance in the input space modified by the correlations lengths. The squared exponential kernel weights this influence as a Gaussian around each training data point.
\begin{figure}[b]
    \centering
    \includegraphics[width=0.99\linewidth]{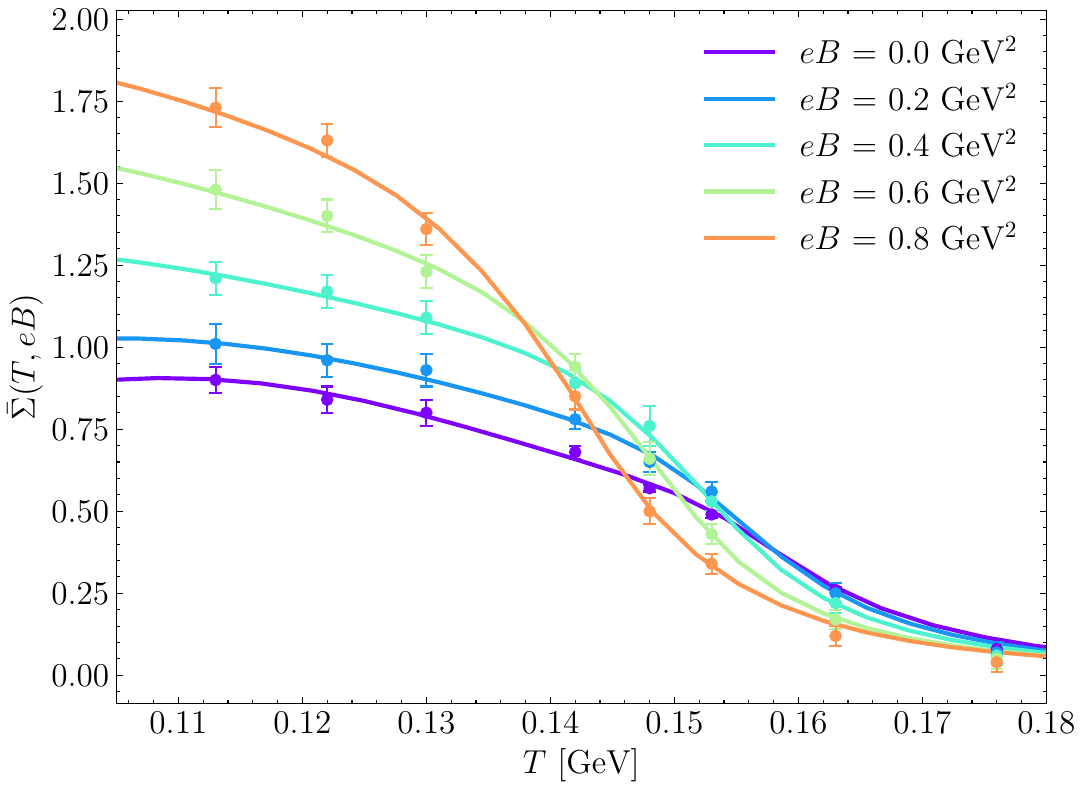}
    \caption{Averaged condensate $\bar\Sigma(T,eB)$ as a function of $T$ for different values of $eB \in \{0.0,0.2,0.4,0.6,0.8\}~\mathrm{GeV}^2$. The solid lines are obtained with the corresponding GP-predicted mean values of $G(T,eB)$, and the data points correspond to lattice results from Ref.~\cite{Bali:2012zg}.}
    \label{fig:Sigma_predicted_values}
\end{figure}
\begin{figure}[t]
    \centering
    \includegraphics[width=0.99\linewidth]{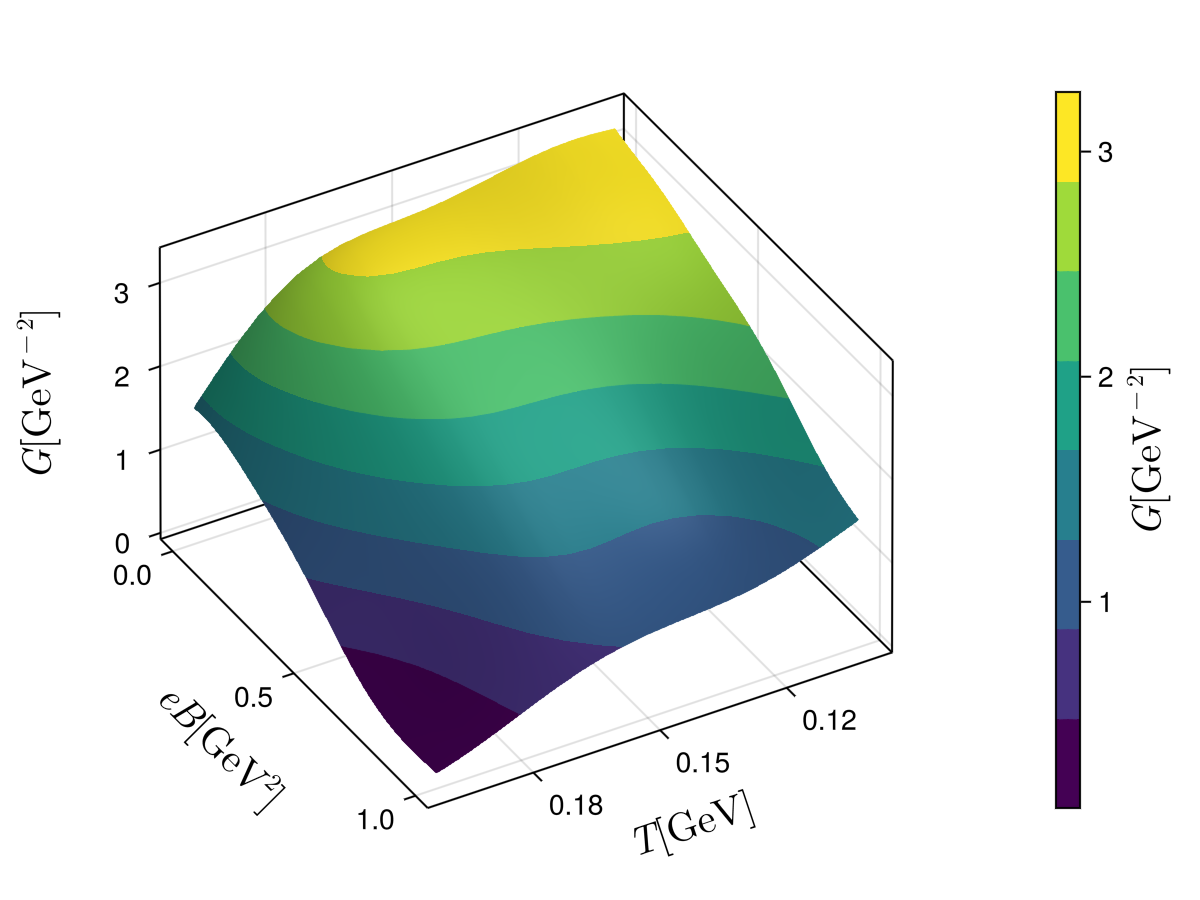}
    \caption{GP-predicted mean values of $G(T,eB)$ as a two-dimensional sheet in the 
    $(T,eB)$ space. The color map classifies the different values of the coupling strength.}
    \label{fig:G_surface}
\end{figure}

As the discrete lattice data for $\bar\Sigma(T,eB)$ was used to fit the continuous $G(T,eB)$, we show in Fig.~\ref{fig:Sigma_predicted_values} the continuous values of $\bar\Sigma(T,eB)$ for different slices of $eB$ where lattice data was available, calculated using the predicted mean $G(T,eB)$ from the GP model. The highly important feature is that the thermomagnetic coupling allows the model to reproduce both the MC and IMC behaviors, as the averaged condensate is enhanced at low temperatures and suppressed near the crossover for increasing magnetic field. This is not the case for fixed coupling $G = G_0$, which only exhibits magnetic catalysis, as may be seen in Ref.~\cite{Farias:2016gmy}. Finally, Fig.~\ref{fig:G_surface} shows the values of the predicted mean of $G(T,eB)$ as a two-dimensional surface in the relevant $(T,eB)$ interval. Throughout most of the range, the coupling decreases with both the temperature and magnetic-field magnitude. Around the chiral crossover, it stays mostly stationary, seemingly indicating that the transition itself is well described by the thermomagnetic PNJL model after $G(T,eB)$ is reduced in the lower temperature region.
%
\section{Numerical Results}~\label{sec:num_results}

In this section, we assess how the thermomagnetic coupling modifies conserved-charge fluctuations and other finite-density observables related to the equation of state. We compare three levels of information: lattice QCD results, the conventional PNJL model with fixed coupling $G_0$, and the thermomagnetic PNJL model constructed from the GP-predicted coupling mean values and their one-standard-deviation interval. We first examine observables at vanishing chemical potential, and then use the Taylor coefficients to reconstruct the pressure and baryon number density at finite $\mu_B/T$. Subsequently, we discuss the magnetic-field dependence of second-order fluctuations, charge correlations, and cumulant ratios.
%
%
\begin{figure*}[t]
\centering
\includegraphics[width=0.325\linewidth]{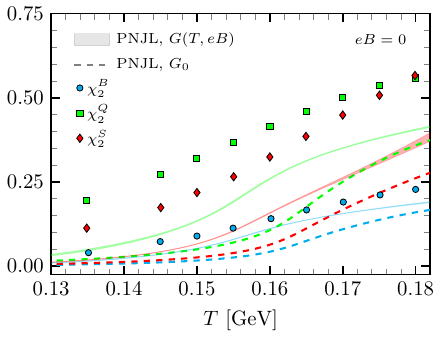}
\includegraphics[width=0.325\linewidth]{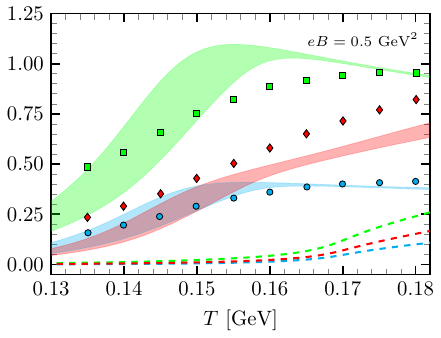}
\includegraphics[width=0.325\linewidth]{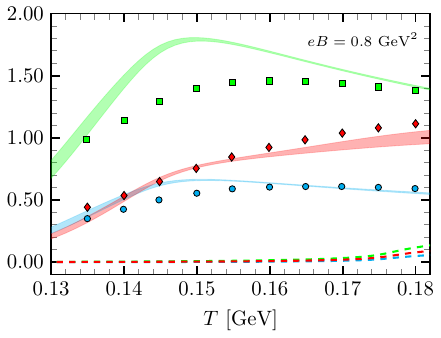}
\caption{Second-order fluctuations of baryon number, electric charge, and strangeness as functions of the temperature for three different values of the magnetic field. Data points correspond to LQCD results from Ref.~\cite{Borsanyi:2023buy}, colored bands to the PNJL model including the medium-dependent coupling $G(T,eB)$, and dashed lines to PNJL with a fixed coupling $G_0$.}
\label{fig:brandt_BBQQSS}
\end{figure*}

To start, we verify the temperature dependence at fixed magnetic field of the diagonal second-order susceptibilities $\chi_{2}^{B}$, $\chi_{2}^{Q}$, and $\chi_{2}^{S}$, which quantify the response of the pressure to baryon-number, electric-charge, and strangeness chemical potentials, respectively. Figure~\ref{fig:brandt_BBQQSS} compares the PNJL results with the lattice QCD data of Ref.~\cite{Borsanyi:2023buy} at $eB=0$, $0.5$, and $0.8~\mathrm{GeV}^{2}$. The colored bands are obtained by propagating the one-standard-deviation confidence interval of the GP reconstruction of $G(T,eB)$. They should therefore be interpreted as a measure of uncertainty associated with the inferred coupling, rather than as a complete estimate of the model uncertainty. The fixed-coupling calculation (dashed lines) remains strongly suppressed throughout the displayed crossover region. This is readily understood as a consequence of its delayed chiral restoration: at a given temperature, the constituent quarks remain too massive and the corresponding charge fluctuations are too small. The issue gets progressively worse as the magnetic field increases, since the $G_0$ case only exhibits magnetic catalysis and the pseudocritical temperature for chiral symmetry restoration grows with $eB$.

Allowing the coupling to decrease with $T$ and $eB$ shifts the melting of the quark condensates to lower temperatures and produces a substantial enhancement of all three number susceptibilities. We must emphasize that the improvement is not due to a direct fit to these observables, since only the light-quark condensate entered the GP determination of $G(T,eB)$ from a different dataset~\cite{Bali:2012zg}, and thus constitutes a nontrivial test of the thermomagnetic parametrization. At vanishing magnetic field, the thermomagnetic model reproduces the monotonic increase of the three observables and their ordering, $\chi_{2}^{Q} > \chi_{2}^{S} > \chi_{2}^{B}$ over most of the interval, but it underestimates their magnitudes. The best agreement is obtained for $\chi_{2}^{B}$, whereas the discrepancy is larger for the charge and strangeness sectors. This flavor dependence is informative, as the coupling was inferred from the average light-quark condensate, while $\chi_{2}^{S}$ is additionally controlled by the slower melting of the strange condensate and $\chi_{2}^{Q}$ is especially sensitive to the unequal electric charges of the quarks.

The magnetic field enhances the number susceptibilities and makes their crossover increasingly pronounced, and the thermomagnetic coupling calculation follows this evolution. At $eB = 0.5~\mathrm{GeV}^{2}$, it describes the lattice baryon fluctuation in close agreement, while at $eB=0.8~\mathrm{GeV}^{2}$ the agreement shifts toward the upper end of the temperature interval. In the latter case, the broad maximum displayed by $\chi_{2}^{B}$ and $\chi_{2}^{Q}$ from the lattice is qualitatively reproduced, although it is also present at $eB = 0.5~\mathrm{GeV}^{2}$ for the model calculation while not clearly defined in the reference data. The full description of $\chi_{2}^{Q}$ remains less accurate, particularly around its maximum, and $\chi_{2}^{S}$ is reproduced only at the qualitative level. The thermomagnetic coupling then captures the magnetic field-driven shift and enhancement of the fluctuations, with the baryon sector described most successfully while residual flavor-dependent discrepancies remain in the charge and strange sectors.
%
%
\begin{figure*}[t]
\centering
\includegraphics[width=0.325\linewidth]{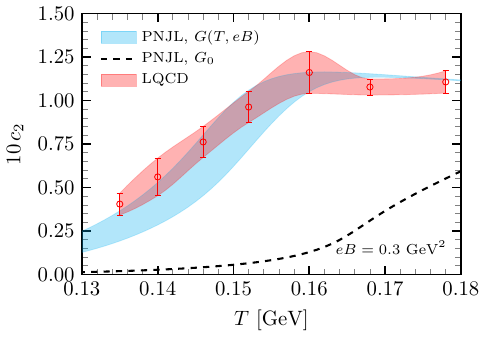}
\includegraphics[width=0.325\linewidth]{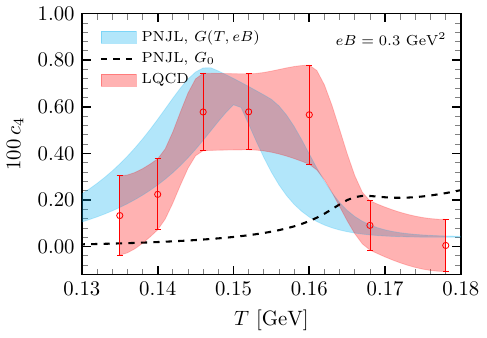}
\includegraphics[width=0.325\linewidth]{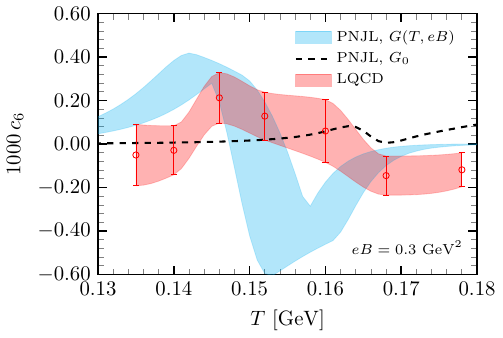} \\
\includegraphics[width=0.325\linewidth]{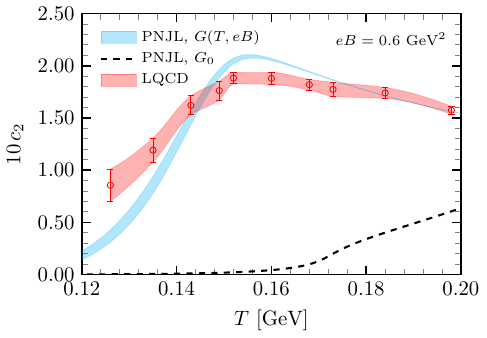}
\includegraphics[width=0.325\linewidth]{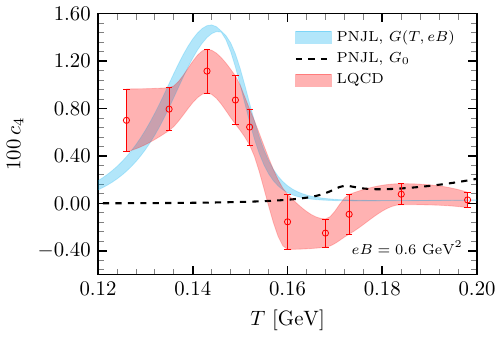}
\includegraphics[width=0.325\linewidth]{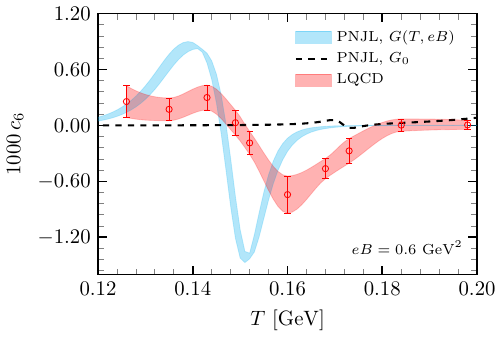}
\caption{Taylor expansion coefficients $c_2$ (left), $c_4$ (middle), and $c_6$ (right) as functions of temperature, for two values of the magnetic field: $eB = 0.3~\mathrm{GeV}^2$ (top) and $eB = 0.6~\mathrm{GeV}^2$ (bottom). Red bands correspond to lattice QCD results from Ref.~\cite{Astrakhantsev:2024mat} in the continuum limit, blue bands to the PNJL model including the medium-dependent coupling $G(T,eB)$, and dashed lines to PNJL with the fixed coupling $G_0$. The coefficients are calculated with the constraints $\mu_u = \mu_d = \mu_B / 3$ and $\mu_s = 0$, approximately corresponding to zero strangeness~\cite{Astrakhantsev:2024mat}.}
\label{fig:Astrakhanstev_cn}
\end{figure*}

Next, we test the response to a nonzero baryon chemical potential through the Taylor coefficients $c_2$, $c_4$, and $c_6$, calculated along the trajectory $\mu_u = \mu_d = \mu_B/3$, $\mu_s = 0$ described in Sec.~\ref{sec:qns_definitions}. Figure~\ref{fig:Astrakhanstev_cn} compares the model with the lattice QCD results of Ref.~\cite{Astrakhantsev:2024mat} for $eB = 0.3$ and $0.6~\mathrm{GeV}^{2}$. Increasing the derivative order progressively emphasizes the narrow crossover region: $c_2$ probes its bulk response, whereas $c_4$ and especially $c_6$ resolve changes in curvature that are much more demanding for an effective mean-field description.

For $c_2$, the thermomagnetic coupling model reproduces the rapid rise toward the crossover and its enhancement with the magnetic field. At $eB = 0.3~\mathrm{GeV}^{2}$, the calculated band approaches a broad maximum around $T \simeq 160~\mathrm{MeV}$ and then decreases mildly, in agreement with the trend of the lattice continuum extrapolation. At $eB = 0.6~\mathrm{GeV}^{2}$, both approaches yield larger values of $c_2$ and a more pronounced decrease after the crossover. However, the model rise is shifted toward slightly higher temperature, and the curvature is not as broad compared to the data. By contrast, the fixed-coupling result remains close to zero over most of the interval, again reflecting a pseudocritical temperature that is too high, and the issue is made worse due to the absence of inverse magnetic catalysis in this case.

The fourth-order coefficient displays a localized peak and is consequently more sensitive to the location and width of the crossover. For $eB = 0.3~\mathrm{GeV}^{2}$ the thermomagnetic coupling model produces a broad maximum near $T \simeq 145$--$150~\mathrm{MeV}$. The lattice maximum occurs at a similar temperature, though the continuum band remains appreciably larger on its high-temperature side. At $eB = 0.6~\mathrm{GeV}^{2}$, the model correctly captures the magnetic enhancement and the sharpening of the positive peak near $145~\mathrm{MeV}$. However, it does not capture the dip toward negative values found in the lattice result. Thus, the thermomagnetic interaction reproduces the overall structure of $c_4$ more successfully than its detailed temperature dependence.

The sixth-order coefficient provides the most stringent comparison. The model yields the expected oscillatory sequence across the crossover---a positive maximum followed by a zero crossing and a negative minimum---whose amplitude grows with $eB$. At $eB = 0.3~\mathrm{GeV}^{2}$, the initial positive structure follows the lattice band with a slight overshoot, but the subsequent negative region is substantially deeper in the model compared to the lattice central values. At $eB = 0.6~\mathrm{GeV}^{2}$, the lattice data clearly favor a positive maximum followed by a negative minimum. The model reproduces this sign pattern, but places both structures at lower temperature and again overestimates the negative amplitude. The larger GP band in this region shows that $c_6$ strongly amplifies the uncertainty in $G(T,eB)$, yet the displacement of the extrema is systematic and cannot be attributed solely to that propagated uncertainty. It points instead to limitations of encoding all medium effects in a single scalar coupling at mean-field level.
\begin{figure*}[t]
\centering
\includegraphics[width=0.45\linewidth]{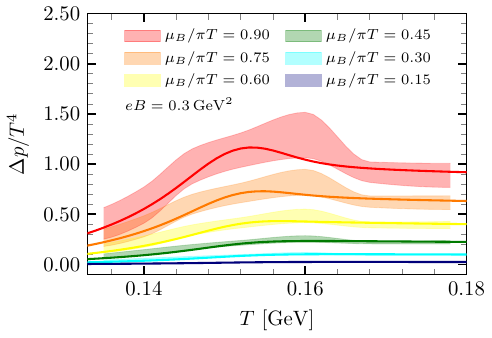}
\includegraphics[width=0.45\linewidth]{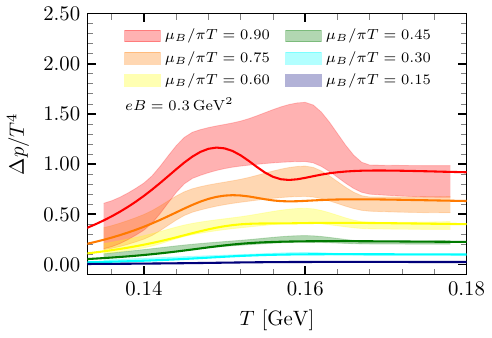}\\
\includegraphics[width=0.45\linewidth]{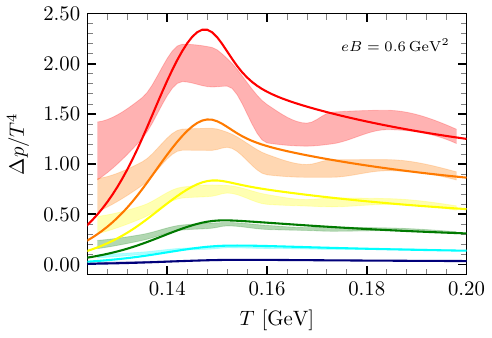}
\includegraphics[width=0.45\linewidth]{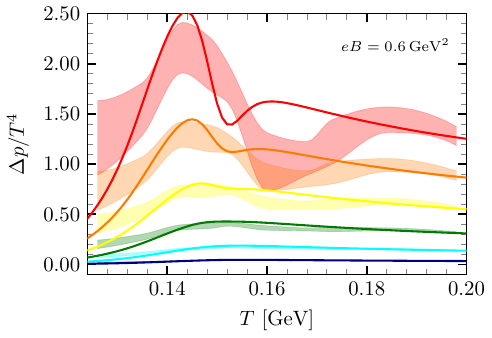}
\caption{Normalized pressure $\Delta p$ as a function of the temperature for different values of the baryon chemical potential ratio $\mu_B / T$, reconstructed using the Taylor expansion coefficients. Left panels correspond to the expansion up to fourth order in $\mu_B / T$, while right panels consider coefficients up to sixth order. Top panels correspond to $eB = 0.3~\mathrm{GeV}^2$, and bottom panels to $eB = 0.6~\mathrm{GeV}^2$. Colored bands represent the LQCD results from Ref.~\cite{Astrakhantsev:2024mat}, and for clarity we show only the PNJL model curves (solid lines) with the mean values of $G(T,eB)$ obtained from the Gaussian process.}
\label{fig:deltap_o4o6}
\end{figure*}

Figure~\ref{fig:deltap_o4o6} shows how these coefficients combine in the finite-density pressure increment $\Delta p/T^{4}$. For visual clarity, only curves with the mean values of $G(T,eB)$ are shown. As required by the Taylor expansion, the signal increases monotonically with $\mu_{B}/T$. At fourth order, the thermomagnetic PNJL curves lie within, or close to, the lattice bands over most of the displayed range. The agreement is particularly good for $\mu_{B}/T \leq 0.6 \pi$, for which the expansion is dominated by $c_2$. Deviations become more visible for $\mu_{B}/T = 0.75 \pi$ and $0.9 \pi$ around the crossover, though the one-standard-deviation contribution to the coupling would give the model more headroom. The stronger enhancement seen at $eB = 0.6~\mathrm{GeV}^{2}$ follows directly from the field dependence of $c_2$ and $c_4$.

Including $c_6$ leaves the low $\mu_{B}/T$ curves almost unchanged, but generates additional peak--dip structures at the two largest chemical potentials. These structures are strongest for $eB = 0.6~\mathrm{GeV}^{2}$, where $c_6$ has its largest amplitude. The sixth-order reconstruction remains broadly compatible with the large lattice uncertainty bands, but it does not systematically improve the comparison over the fourth-order result. This order-by-order behavior is important, as it indicates that the bulk pressure is robust at moderate $\mu_{B}/T$, whereas its fine structure near the crossover is already sensitive to the least accurately reproduced coefficient.
\begin{figure*}[t]
\centering
\includegraphics[width=0.45\linewidth]{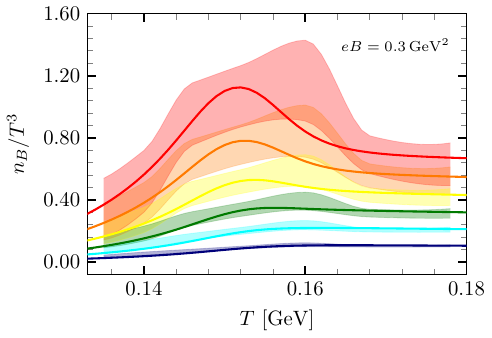}
\includegraphics[width=0.45\linewidth]{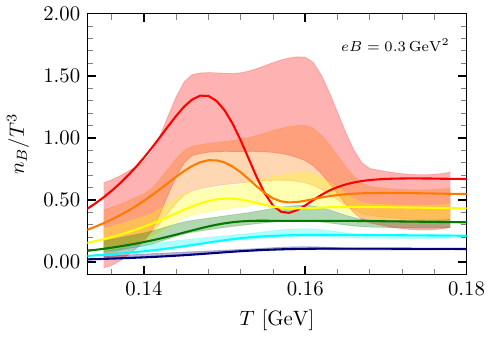} \\
\includegraphics[width=0.45\linewidth]{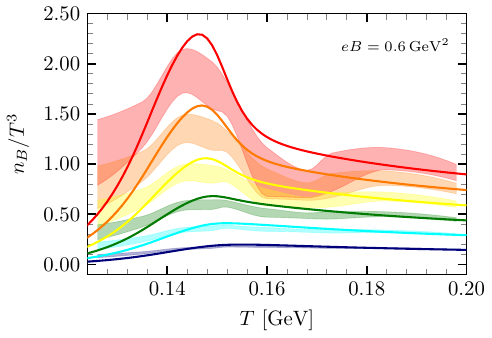}
\includegraphics[width=0.45\linewidth]{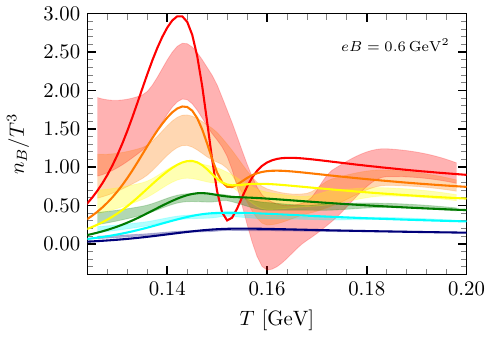}
\caption{Baryon density $n_B$ as a function of the temperature, following the same panel structure as Fig.~\ref{fig:deltap_o4o6}. LQCD data from Ref.~\cite{Astrakhantsev:2024mat}.}
\label{fig:nb_o4o6}
\end{figure*}

The same comparison for the baryon density is presented in Fig.~\ref{fig:nb_o4o6}. Because $n_{B}$ is obtained by differentiating the pressure with respect to $\mu_{B}$, the relative contribution of each higher-order coefficient is enhanced. At fourth order, the model follows the lattice ordering with $\mu_{B}/T$ and reproduces the broad crossover maximum, although the model peak is generally narrower and occurs at a slightly lower temperature. The overall magnitude and magnetic enhancement are reasonably described, particularly for the smaller values of $\mu_{B}/T$.

The sixth-order term has a much larger visible impact on $n_{B}$ than on $\Delta p$. For $\mu_{B}/T = 0.9 \pi$, it produces a pronounced suppression around $T \simeq 155~\mathrm{MeV}$ at $eB = 0.3~\mathrm{GeV}^{2}$ and even a narrow negative digression at $eB = 0.6~\mathrm{GeV}^{2}$ for the lattice band. The crossing of curves near the crossover for high $\mu_{B}/T$ ratios is also an artifact due to uncertainties on $c_6$. A negative baryon density at positive $\mu_{B}$ is not interpreted as a physical prediction. Rather, it signals that the sixth-order truncated series is locally unreliable when the value of $c_6$ is large and negative. The effect rapidly decreases with smaller $\mu_{B}/T$, supporting the use of the reconstruction only in the range where successive truncation orders remain stable.
%
%
\begin{figure*}[ht]
\centering
\includegraphics[width=0.45\linewidth]{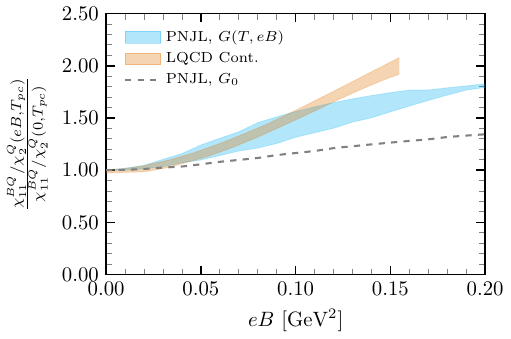}
\includegraphics[width=0.45\linewidth]{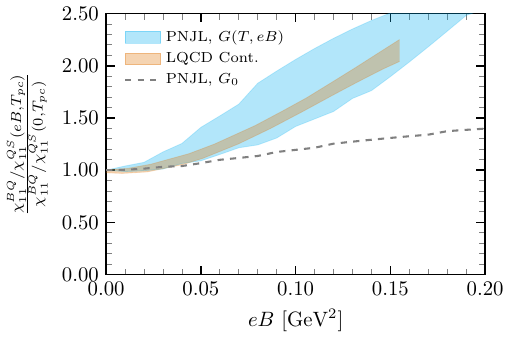} \\
\includegraphics[width=0.45\linewidth]{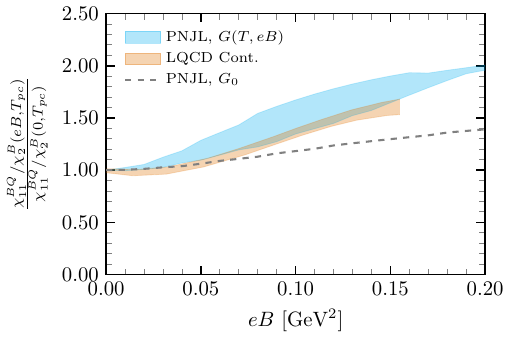}
\includegraphics[width=0.45\linewidth]{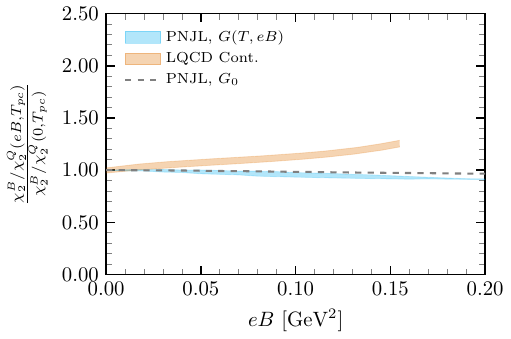}
\caption{Ratios of second order correlations and fluctuations as functions of the magnetic field, calculated along the pseudocritical temperature $T_{pc}$. Brown bands correspond to LQCD results from Ref.~\cite{Ding:2025jfz}, blue bands to the PNJL model including the medium-dependent coupling $G(T,eB)$, and dashed lines to PNJL with a fixed coupling $G_0$. All ratios are normalized by their corresponding zero-field values.}
\label{fig:comp_fig8}
\end{figure*}

We finally turn to the magnetic-field dependence of charge correlations. Ratios along the pseudocritical line are shown in Fig.~\ref{fig:comp_fig8}, in comparison to continuum-extrapolated lattice data from Ref.~\cite{Ding:2025jfz}. The normalization to the zero-field value reduces the sensitivity to the absolute scale of each observable and isolates its magnetic response. The lattice results show a substantial enhancement of the first three ratios over the range $eB \lesssim 0.16~\mathrm{GeV}^{2}$. The thermomagnetic model reproduces this monotonic increase and its approximate magnitude. The fixed-coupling calculation, in contrast, gives a significantly weaker response, demonstrating that Landau-level kinematics alone is insufficient and that the inclusion of inverse magnetic catalysis is essential.

The lower-right ratio behaves differently. The lattice band rises mildly, whereas both the thermomagnetic and constant coupling results remain nearly constant or decrease slightly. Because this ratio combines channels with different flavor weights, part of the common magnetic enhancement cancels, making it particularly sensitive to the relative light- and strange-sector responses. The residual discrepancy is therefore consistent with the limitations already observed for $\chi_{2}^{S}$ in Fig.~\ref{fig:brandt_BBQQSS}. Overall, the cumulant ratio $\sim \chi_{11}^{BQ} / \chi_{11}^{QS}$ shows the largest enhancement with increases in the magnetic field magnitude, which we elaborate on in sequence. To clarify, small irregularities in the blue bands should not be interpreted as physical oscillations here. They reflect the discrete determination of $T_{pc}(eB)$ and the numerical interpolation of the GP coupling along that trajectory.
\begin{figure*}[ht]
\centering
\includegraphics[width=0.325\linewidth]{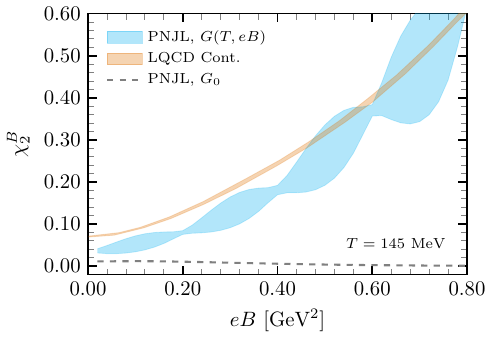}
\includegraphics[width=0.325\linewidth]{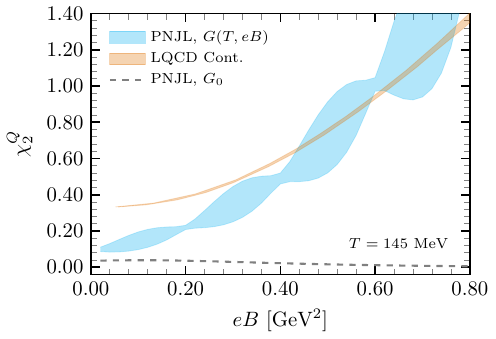}
\includegraphics[width=0.325\linewidth]{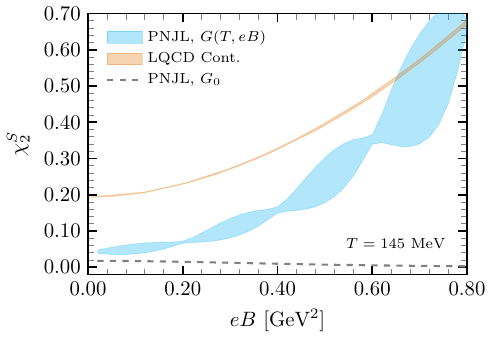} \\
\includegraphics[width=0.325\linewidth]{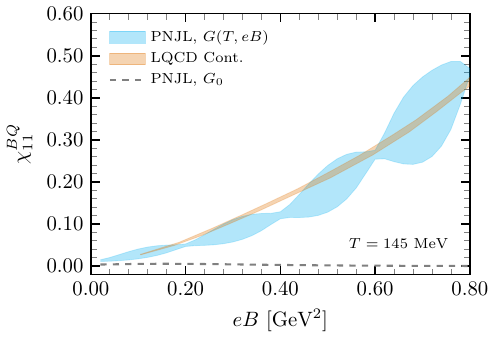}
\includegraphics[width=0.325\linewidth]{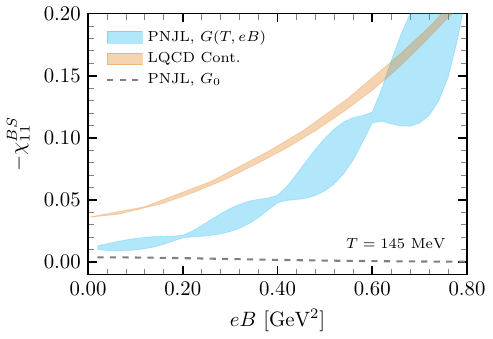}
\includegraphics[width=0.325\linewidth]{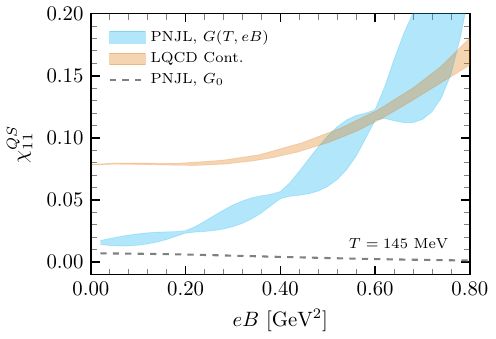}
\caption{Second-order correlations and fluctuations as functions of an extended range of the magnetic field, calculated at fixed temperature $T = 145~\mathrm{MeV}$. As before, brown bands correspond to LQCD results from Ref.~\cite{Ding:2025jfz}, blue bands to the PNJL model including the medium-dependent coupling $G(T,eB)$, and dashed lines to PNJL with a fixed coupling $G_0$.}
\label{fig:comp_fig9}
\end{figure*}

Figure~\ref{fig:comp_fig9} extends the analysis to $eB = 0.8~\mathrm{GeV}^{2}$ at the fixed temperature $T = 145~\mathrm{MeV}$, near but slightly before the crossover. All six diagonal and off-diagonal observables are enhanced by the magnetic field, but with clearly different rates. The strongest growth occurs in the electric-charge fluctuation, as expected from the direct coupling of the Landau spectrum to the quark electric charges. The baryon and strangeness fluctuations grow more moderately, while the mixed $BQ$, $BS$, and $QS$ channels track the corresponding flavor correlations. In relative terms, however, $\chi_{11}^{BQ}$ shows the greatest enhancement with the magnetic field compared to its $eB = 0$ value. This fractional response is the physically measurable quantity, which prompted the proposal of $\chi_{11}^{BQ}$ as the magnetometer of QCD by Refs.~\cite{Ding:2023bft,Ding:2025jfz}. Noticeably, it is also the correlation that shows the best agreement between lattice data and our PNJL model framework.

Overall, the thermomagnetic coupling calculation captures the strong nonlinear increase and approaches the lattice bands at intermediate and large fields for most channels. At small $eB$, however, it tends to underestimate the absolute magnitude, most visibly in the strange and mixed strange-charge sectors. The nearly vanishing fixed-coupling curves over the same interval emphasize once more that its crossover lies far beyond the thermodynamic region probed at $T = 145~\mathrm{MeV}$. The waviness of the thermomagnetic bands follows the local GP reconstruction and should be regarded as interpolation sensitivity rather than evidence for resolved oscillatory physics. The uncertainty is smaller precisely at $eB = \{0.0, 0.2, 0.4, 0.6, 0.8\}~\mathrm{GeV}^{2}$, the points on which the light-quark condensate constraints were imposed. Taken together, Figs.~\ref{fig:comp_fig8} and~\ref{fig:comp_fig9} show that the coupling inferred solely from the condensate provides substantial predictive information to independent conserved-charge observables, while the remaining channel dependence identifies where additional flavor-dependent interactions or constraints would be most valuable.
%
\section{Conclusions}~\label{sec:conclusions}

We have investigated the equation of state and conserved-charge fluctuations of magnetized QCD matter at finite temperature and chemical potential within a three-flavor PNJL model. The central ingredient is a thermomagnetic four-quark interaction coupling, $G(T,eB)$, inferred from lattice QCD results for the average light-quark condensate through a Gaussian Process regression. This construction incorporates inverse magnetic catalysis directly into the effective interaction without imposing a fixed functional form on its temperature and magnetic-field dependence.

A fundamental result is that information extracted from a single chiral observable is transferred to a broad set of independent thermodynamic quantities. Relative to the fixed-coupling PNJL model, the thermomagnetic interaction substantially improves the temperature and magnetic-field dependence of the diagonal second-order susceptibilities. The improvement is most convincing in the baryon sector. Electric-charge and strangeness fluctuations exhibit the correct qualitative enhancement, but retain channel-dependent discrepancies in magnitude and around their field-induced maxima. The GP bands quantify only the uncertainty propagated from $G(T,eB)$ and should not be interpreted as a complete estimate of the systematic uncertainty of the PNJL framework.

The Taylor coefficients provide a progressively more sensitive test of the model. The thermomagnetic PNJL calculation reproduces the rise and magnetic enhancement of $c_2$, as well as the localized crossover structure of $c_4$, although the positions and widths of the extrema are not reproduced uniformly. For $c_6$, the model captures the alternating positive--negative pattern seen across the crossover, but places the extrema at lower temperatures and overestimates the negative digression. These deviations show that fitting the light-quark condensate constrains the bulk thermodynamic response more effectively than the higher derivatives that resolve the detailed crossover structure.

The finite-density reconstructions reveal a corresponding hierarchy of stability. The pressure is comparatively insensitive to the inclusion of the sixth-order term at low and moderate $\mu_B/T$, whereas the baryon density amplifies the uncertainty and sign structure of $c_6$. At the largest $\mu_B/T$ ratios considered, the sixth-order PNJL reconstruction develops an inverse behavior near the crossover, where the magnitude of $n_B$ decreases for higher values of the scaled baryon chemical potential. We do not interpret these regions as physical phases. This type of response instead signals a local loss of reliability of the truncated series, and is also visible in the lattice bands. The continuation from $\mu_B = 0$ is predictive only where successive truncation orders remain stable, and the thermodynamic observables preserve their required positivity properties.

The diagonal and mixed second-order correlations further demonstrate that the magnetic response cannot be described by Landau-level kinematics alone. Both along the pseudocritical line and at a fixed temperature slightly before it, the thermomagnetic coupling produces the strong nonlinear enhancement observed in lattice QCD, whereas the fixed-coupling calculation remains strongly suppressed. However, the agreement is again nonuniform, as ratios dominated by the light-quark response are described more successfully than combinations in which light- and strange-sector contributions partially cancel. These observables therefore provide useful flavor-dependent constraints beyond those contained in the average light-quark condensate.

Altogether, these results identify the medium dependence of the effective quark interaction as a key ingredient missing from the conventional PNJL description of magnetized matter. A coupling inferred solely from the chiral condensate accounts for a substantial fraction of the discrepancy with independent lattice observables and provides a noticeably improved baseline for the equation of state at nonzero magnetic field. The comparison does not, however, support a uniform quantitative agreement across all channels or derivative orders. The remaining discrepancies expose the limitations of encoding the full QCD medium dependence in a single flavor-independent scalar coupling at mean-field level. Higher-order fluctuations are sensitive to the detailed chiral crossover, the Polyakov-loop potential, flavor-dependent interactions, and correlations beyond mean field. The strange and mixed-charge channels additionally suggest that constraining only the average light-quark condensate is insufficient for a fully quantitative description.

Within these limitations, the present framework provides a direct bridge between first-principles information at $\mu_B = 0$ and an effective description of magnetized matter at nonzero baryon density. More broadly, our results support lattice-constrained, medium-dependent interactions as a promising strategy for extending effective QCD models into regions that remain difficult to access directly from first principles. A natural next step is to perform a multi-observable inference in which the four- and six-quark interactions along with the Polyakov-loop sector are constrained simultaneously, including flavor-sensitive lattice information. Additional interaction channels, in particular vector interactions, may also modify the finite-density response and thus provide another path for future investigations.
%
%
\begin{acknowledgments}
This work was partially supported by Conselho Nacional de Desenvolvimento Cient\'ifico e Tecno\-l\'o\-gico  (CNPq), Grants No. 312032/2023-4, 402963/2024-5 and 445182/2024-5 (R.L.S.F.), No. 141270/2023-3 and 201300/2025-7 (B.S.L.), No. 176811/2026-5 (J.J.B.F.), and No. 311199/2026-7 (D.C.D.); Funda\c{c}\~ao de Amparo \`a Pesquisa do Estado do Rio 
Grande do Sul (FAPERGS), Grants No. 24/2551-0001285-0 (R.L.S.F.) and 23/2551-0001591-9 (D.C.D.); CAPES Finance Code 001 (A.E.B.P.); The work is also part of the project
Instituto Nacional de Ci\^encia e Tecnologia - F\'isica Nuclear e
Aplica\c{c}\~oes (INCT - FNA), Grants No. 464898/2014-5 and 408419/2024-5, and supported
by the Ser\-ra\-pi\-lhei\-ra Institute (grant number Serra -
2211-42230). B.S.L. and R.L.S.F. acknowledge
the kind hospitality of the Center for Nuclear Research at Kent State University, where part of this work was
done.
\end{acknowledgments}

\bigskip \noindent {\it Data Availability.} The data are not publicly available. The data are available from the authors upon reasonable request.
%
%
\bibliography{references}
\end{document}